\documentstyle[aps,prl,epsfig,multicol,fancyheadings,amsbsy,amssymb,amstex]{revtex}
\begin{document}
\title{Pairing Correlations in a Generalized Hubbard Model 
for the Cuprates.}
\author{Liliana Arrachea$^{a}$ and A.A. Aligia$^{b}$.}
\address{$^{a}$ Departamento de F\'{\i }sica, \\
FCyN Universidad de Buenos Aires\\
Pabell\'{o}n I, Ciudad Universitaria, (1428) Buenos Aires, Argentina.\\
$^{b}$ Centro At\'{o}mico Bariloche and Instituto Balseiro,\\
Comisi\'on Nacional de Energ\'{\i}a At\'{o}mica, \\
8400 Bariloche, Argentina.}
\date{Received \today }

\maketitle

\begin{abstract}
Using numerical diagonalization of a 4x4 cluster, we calculate  
on-site $s$, extended $s$ and $d_{x^{2}-y^{2}}$ pairing correlation 
functions (PCF) in an effective generalized Hubbard model 
for the cuprates,
with 
nearest-neighbor correlated hopping  and next 
nearest-neighbor hopping $t'$.
The vertex contributions (VC) to the PCF are significantly 
enhanced, relative to the $t-t^{\prime}-U$ model. The behavior of
the PCF and their VC,
 and signatures of anomalous flux quantization, indicate 
superconductivity in the $d$-wave channel for moderate doping
and in the $s$-wave channel for high doping 
and small $U$.
\end{abstract}


\vspace*{-1.1cm}
\begin{multicols}{2}

\columnseprule 0pt

\narrowtext

Since the discovery of high temperature superconductivity, much effort has
been devoted to study the properties of the Hubbard model, the $t-J$ model,
and modifications of them. While these studies have helped to clarify
several optical and magnetic properties of the cuprates \cite{dag0,kam}, 
the superconducting mechanism 
remains unclear. Studies in
generalized $t-J$ models suggest a magnetic origin of superconductivity 
\cite{dag,rvb,pla,fei,tsu}, but the numerical results seem to require 
either a
superexchange $J$, or a three-site term \cite{rvb,tsu}, which is beyond the
realistic range for the cuprates. In addition, the constraint of no double
occupancy in these models reduces the mobility of the superconducting
 pairs 
\cite{jec}. On the other hand, the search for signals of superconductivity
in the Hubbard model have been negative so far \cite{dag0,mor,zha}. This
fact stimulates the study of modifications of the Hubbard model which
represent more closely the physics of the cuprates \cite{jec,dwave}.
Recently an effective modified Hubbard model for the cuprates derived
earlier \cite{sch} has been studied \cite{dwave}. The model includes a
nearest-neighbor (NN) correlated hopping which depends on the 
occupation of the two
sites involved and next-NN hopping $t^{\prime }$.
Within a mean-field approximation \cite{dwave},
the correlated hopping has been found to originate pairing, the
 underlying mechanism being similar to 
that provided by a superexchange coupling $J$ \cite{jec,dwave,1d}.
The shape of the Fermi surface and the positions of the van Hove
singularities (vHS), modified with $t^{\prime}$,
influence the magnitude and the symmetry of 
the order parameter.
The  expected instability for moderate dopings,
is $d$-wave superconductivity
in concurrence with long-range antiferromagnetism near 
half-filling.

In this Letter we report results on pairing correlation functions (PCF) and
spin correlation functions for this effective model, obtained 
by numerical diagonalization of a square cluster containing 
$L=16$ unit cells. We find evidence 
of strong superconducting correlations with $d_{x^2-y^2}$-wave 
symmetry in the doping regime of interest for the cuprates. 
Furthermore, in contrast to the case of the ordinary
Hubbard model, we find indications of anomalous flux quantization (AFQ),
characteristic of superconductivity \cite{afq}, in most of the explored
region of parameters.
 Our numerical results support the mean-field picture. 
The size of the cluster and
the inclusion of $t^{\prime }$ made the calculation
 particularly difficult. To our
knowledge, even with $t^{\prime }=0$, no exact PCF have been so far reported
in this cluster allowing doubly occupied sites.  

The effective model for the cuprates is \cite{sch}: 
\begin{eqnarray}
H &=&U\sum_{i}n_{i\uparrow }n_{i\downarrow }\;-\;t^{\prime
}\sum_{<ij^{\prime }>\sigma }\;c_{i\sigma }^{\dagger }c_{j^{\prime }\sigma }
\nonumber \\
&-&\sum_{<ij>\sigma }(c_{i\bar{\sigma}}^{\dagger }c_{j\bar{\sigma}%
}+h.c)\{t_{AA}\;(1-n_{i\sigma })(1-n_{j\sigma })+ \nonumber \\
& & t_{BB}n_{i\sigma
}n_{j\sigma } 
+t_{AB}[n_{i\sigma }(1-n_{j\sigma })+n_{j\sigma }(1-n_{i\sigma })]\},
\label{ham}
\end{eqnarray}
where $<ij>$ ($<ij^{^{\prime }}>$) denotes NN (next-NN) positions of the
lattice. $U$ represents the cost in energy of constructing a Zhang-Rice
singlet from two singly occupied cells. $t_{AA}$ represents the hopping of a
Zhang-Rice singlet to a singly occupied NN cell. The terms
with amplitude $t_{AB}$ correspond to the destruction of a Zhang-Rice singlet
and a nearest-neighbor cell without holes, creating two singly occupied
cells and vice versa. $t_{BB}$ describes the movement of an isolated hole.
While $U$ lies between 3 and 4eV, the magnitude of the correlated hopping
terms is ten times smaller, and $t_{AB}$ $\sim 10\%$ larger than 
$(t_{AA}+t_{BB})/2$ has been estimated \cite{sch}. However for other
parameters of the multiband model, this ratio can be much larger, since 
$t_{AB}$ is linear in the Cu-O hopping $t_{pd}$, 
while $t_{AA},t_{BB}  \sim $ $t_{pd}^{2}$ \cite{dwave}.
In mean field, for $t_{AB} > t_{AA}, t_{BB}$,
superconductivity in the $s$- and $d$-wave channels is obtained \cite{dwave}.
Near half filling, $d$-wave superconductivity
 competes with long-range
antiferromagnetism. If $t^{\prime }=0$, a SDW takes place at $n=1$ while 
finite $t^{\prime }$
destroys perfect nesting, and for doping such that vHS
 lie near the Fermi level, a $d$-wave superconductor
coexisting with short-range antiferromagnetic fluctuations is expected. 
Instead,
$s$-wave superconductivity develops for small $U$ and sufficiently small
particle densities $n$. 
Although vHS are not well defined in a small
cluster, $t^{\prime}$ introduces changes in the distribution
of the particles in $k$-space and
conclusions concerning the
tendencies in the behavior of the PCF can be extracted.
We restrict to the electron-hole symmetric
case $t_{AA}=t_{BB}=1$ and large $t_{AB} \geq 2$, in
order to render more noticeably the effects of the correlated hopping. We
also investigate $t^{\prime }=0,-0.45$ \cite{tpri}.

The PCF are: 
\begin{equation}
P_{\alpha }(i)\;=\;
\langle \Delta _{\alpha }^{\dagger }(i)\Delta _{\alpha}(0)\rangle ,  
\label{pcf}
\end{equation}
where for on-site $s$ pairing $\Delta _{os}^{\dagger }(i)=c_{i\uparrow
}^{\dagger }c_{i\downarrow }^{\dagger }$, while $\Delta _{\alpha }^{\dagger
}(i)=\sum_{\delta }f_{\alpha }(\delta )[c_{i+\delta \uparrow }^{\dagger
}c_{i\downarrow }^{\dagger }-c_{i+\delta \downarrow }^{\dagger }c_{i\uparrow
}^{\dagger }]/\sqrt{8}$, with $f_{es}(\delta )=1$ for extended $s$ pairing,
and $f_{d}(\delta )=1$ ($f_{d}(\delta )=-1$ ) when $\delta =\pm (1,0)$ 
($\delta =\pm (0,1)$) for $d_{x^{2}-y^{2}}$ pairing. We normalize $\Delta
_{\alpha }^{\dagger }(i)$ in such a way that $|\Delta _{\alpha }^{\dagger
}(i)|0\rangle |^{2}=1$, to facilitate comparison among the different PCF 
\cite{note}. To compute the
VC to the PCF \cite{whi}, 
denoted as $\bar{P}_{\alpha }(i)$,  
the quantity $(\langle c_{\lambda }^{\dagger }c_{\xi }\rangle
\langle c_{\mu }^{\dagger }c_{\nu }\rangle -\langle c_{\lambda }^{\dagger
}c_{\nu }\rangle \langle c_{\mu }^{\dagger }c_{\xi }\rangle )$ is
subtracted,
for every term in Eq. (\ref{pcf}) of
the form $\langle c_{\lambda }^{\dagger }c_{\mu }^{\dagger }c_{\nu }
c_{\xi}\rangle $. For a BCS ground state, $\bar{P}_{\alpha }(i)$ is 
positive and
proportional to the square of the order parameter.
The results we show for the correlations functions correspond to
``optimum'' boundary conditions (OBC), which could be
periodic (PBC), antiperiodic (ABC) or
mixed (MBC), i.e. periodic in one direction and antiperiodic in the other, 
according to those which lead to the minimum ground-state energy.
The computation has been
made possible by exploiting  
all symmetry operations of the
space group of the square lattice \cite{fano} plus time reversal (256
operations in the cluster). 
Half of these operations are lost for MBC, and the reported PCF
are averages over equivalent distances in the periodic system.

To give an idea of the expected magnitude of the PCF
and in order to establish a criterion to interpret our
results, we analyze the behavior of
the PCF and the VC for the usual attractive Hubbard model 
with a quite large attraction $U=-5$, in which case 
superconductivity is 
 well supported by several calculations \cite{negu}.
These quantities are displayed in Fig. 1 for
$N=10$ particles and distances larger than one lattice site
\cite{rie}.   
As in the case of previous Monte Carlo results \cite{zha}, 
$P_{\alpha }(r)$ shows oscillations with distance $r$, while 
$\bar{P}_{\alpha }(r)$ exhibits a smoother behavior. 
It is clear that  $\bar{P}_{os}(r)$ dominates over the other 
PCF, which is in agreement with the $s$-wave, predominantly on-site
character of the superconductivity in the model \cite{negu}.
In the light of these results, we establish the following criterion
to extract information from our numerical data:
we conclude that superconducting correlations in the $\alpha$-channel 
are present in the model when {\em both quantities}, 
$P_{\alpha}$ and $\bar{P}_{\alpha}$, are enhanced at large 
distances relative to the non-interacting case.

In Fig. 2, we show the effect of $t_{AB}$ and $t^{\prime }$ 
for $U=0$. 
The PCF $P_{\alpha }(r)$ (not shown) display the same
qualitative behavior as those in Fig. 1 (a).  
We conclude that for these parameters the model has strong
signals of $s$-wave superconductivity in both on-site and NN channels.
This agrees with the mean-field calculations \cite{dwave}. For $N=12$
particles, the values of $\bar{P}_{\alpha }(r)$ (not shown) are reduced 
in $\sim 0.01$, but the qualitative behavior remains the same. For 
$t^{\prime }=0$, superconducting correlations are also strongest in
 the $s$-channel. For $N=12$,  $\bar{P}_{es}(\sqrt{8})$ is 
approximately one half of the corresponding value for $N=10$.
For both densities, a negative $t'$ enhances  $\bar{P}_{es, os}(r)$ 
relative to the case with  $t'=0$. 
According to Ref. \cite{dwave}, when $U$ overcomes a certain value, 
$s$-wave superconductivity 
is replaced by a SDW when $t^{\prime}=0$ and by
$d$-wave superconductivity for finite $t^{\prime}$. 
Keeping $t^{\prime }=-0.45$, $t_{AB}=2$, and
increasing $U$, we find a decrease in the $s$-wave 
PCF and
an increase in the $d$-wave ones. 
The latter dominate already for $U=4$ and $N=10$, 
 with $\bar{P}_{d}(r)\sim 0.015$
and values significantly larger than those for $t_{AB}=1$.

For $U=10$, with $t_{AB}=2$, the PCF are much larger in the $d$-wave channel. 
The behavior of $P_d(r)$ and  $\bar{P}_d(r)$ for different densities
($N/L$=0.625, 0.75 and 0.875)
is shown in Fig. 3, for $t^{\prime}=0, -0.45$. 
To simplify the figure, we do not show the values of $P_d(r)$ 
for $t_{AB}=1$ and for the non-interacting case. For $N=10, 12$,
with $t_{AB}=2$  and $t'=-0.45$,
the values $\bar{P}_{d}(r) \sim 0.02, 0.03$
(Fig. 3(d)) at distances $\sqrt{2}\leq r\leq \sqrt{8}$ 
are roughly half of the values of 
$\bar{P}_{os}(r)$ for the Hubbard model with strong on-site attraction 
(Fig. 1), 
and very similar to those of $P_{d}(r)$ for a short-range 
resonance-valence-bond wave
function which by construction has superconducting ODLRO \cite{rvb,note}.
These results are strong indications of $d$-wave superconductivity. 
We should also note that the superconducting $d$-wave pairs in the model,
have an internal structure which extends beyond NN and with only a
partial overlap with $\Delta _{d}^{\dagger }(i)$. Thus, our $d$-wave PCF are
reduced with respect to the optimum normalized PCF by the square of this
overlap \cite{jec}. 

For $N=10, 12$, the effect of a negative 
$t^{\prime}$ is to enhance the VC $\bar{P}_d(r)$. 
Instead, for $N=14$, both $P_d(r)$ and  $\bar{P}_d(r)$ are large for
the case with $t^{\prime}=0$, while they are very small for 
$t^{\prime}=-0.45$.  
Note that in all the  cases with sizable pairing correlations, 
the values of $\bar{P}_d(r)$ corresponding to $t_{AB}=2$ are significantly
larger than those corresponding to $t_{AB}=1$, with the same values of
$t^{\prime}$ and $U$. In addition, in these cases, the non-interacting
$P_d(r)$ lie bellow the displayed ones for $t_{AB}=2$ in Fig. 3
(a). The remarkable large values of $\bar{P}_d(r)$
observed in Fig. 3(c) for the case with $N=14$ particles could 
be somewhat exaggerated due to particular finite-size effects \cite{notebc}.
In fact, when $t_{AB}=1$, 
$\bar{P}_d(r)$ in Fig. 3 (c) is large while the values of
$P_d(r)$ are smaller than those of the non-interacting case. 
The mean-field
treatment \cite{dwave} predicts a maximum of the superconducting gap
with  $d$-wave symmetry at half-filling for $t^{\prime}=0$, 
when long-range antiferromagnetism is not taken into account.
The 
concurrence between superconductivity and long-range 
antiferromagnetism near half-filling
manifests itself in this cluster  when different BC are used.
For $N=14, t^{\prime}=0, t_{AB}=2, U=10$, 
spin-spin correlations (not shown)
are much stronger for PBC than for MBC,
while 
in the first (latter) case $P_d(r)$ is weaker (stronger)
than in the non-interacting case.
In any case, as expected \cite{fei,tsu,dwave}, the maximum 
of the PCF with doping shifts to higher doping as $t'$ increases.

In contrast to the cases without correlated
hopping, we find signs of AFQ in most of the explored parameter space.
AFQ consists of a periodicity of half a flux quantum in the ground-state
energy $E(\Phi )$ as a function of a flux $\Phi $ threading the system in a
toroidal geometry, and it is a necessary but not sufficient condition for
superconductivity \cite{afq}. In finite systems, a tendency to AFQ is
indicated by a crossing of energy levels with different total wave vector as 
$\Phi $ is varied, and the presence of two relative minima in $E(\Phi )$
with a difference of $\Phi $ in $\pi $ (usually at $\Phi =0$ and 
$\Phi $=$\pi $) in the interval $[0,2\pi )$. 
In Fig. 4 we show the dependence on $\Phi $ of the lowest energy levels 
of the system for several values of the
parameters in which the level crossing occurs. Fig. 4 (a) and (b)
correspond to dominant $s$-wave PCF. In the other cases shown, the $d$-wave
PCF are the largest ones. 
For $t^{\prime}=0, N=10$ we found a very similar behavior of
$E(\Phi )$ to that observed in Fig. 4.
Due to the fact that the introduction of a flux breaks
the space group symmetry and increases considerably the size of the
irreducible subspaces, we have not constructed $E(\Phi )$ curves for $N>10$.

We have also calculated charge and spin correlation functions. The spin
structure factor 
$S({\bf q})=\sum_{ij}\langle S_{i}^{z}S_{j}^{z}e^{i{\bf q}
({\bf R}_{i}-{\bf R}_{j})}\rangle /L^{2}$ 
for $N=12$ is shown in Fig. 5. 
In all cases, the increase of $t_{AB}$
tends to restore a peak at $(\pi , \pi )$, which is rather broad,
indicating the presence of short-range antiferromagnetism, 
similar to that found for an  RVB state with superconducting
ODLRO \cite{rvb}. 
For the case $N=14, t^{\prime}=0, t_{AB}=2, U=10$  $S({\bf q})$
exhibits a broad structure at $(\pi , \pi )$ for MBC, while large 
$P_d(r)$ is obtained, as discussed above.
Instead, for PBC a much narrower peak, suggestive of longer-range
antiferromagnetic correlations is observed in $S({\bf q})$
while $P_d(r)$ are weaker than those of the non-interactive case. 
In one dimension (1D), there is analytical and numerical evidence that for
small $U$ and large $t_{AB}$, the ground state at half filling consists of
singlet dimers, and singlet PCF dominate when the system is doped \cite{1d}.
The natural extension of this scenario to 2D, seems to be
a short range RVB-like state
 at half filling, which turns into a singlet superconductor as the dimers 
acquire
mobility with doping. 
While we expect long-range antiferromagnetism in the half-filled case,
our results for PCF and $S({\bf q})$ are consistent
with this scenario
as the system is doped.

In summary, we have shown that correlated hopping, which arises naturally in
a low-energy reduction of multiband models for the cuprates, leads to
pairing correlations of magnitude similar to that observed in
the same cluster for models for which superconductivity is well
established. For low to moderate doping, 
the favored symmetry is  $d_{x^{2}-y^{2}}$ for finite $U$ and 
$t^{\prime }$. Part of the numerical work was done at the Max-Planck
 Institut PKS. L. A
acknowledges support from CONICET. A. A. A. is partially supported by
CONICET. This work was sponsored by PICT 03-00121-02153 of ANPCyT and PIP
4952/96 of CONICET.

\begin{figure}[htbp]
\epsfig{file=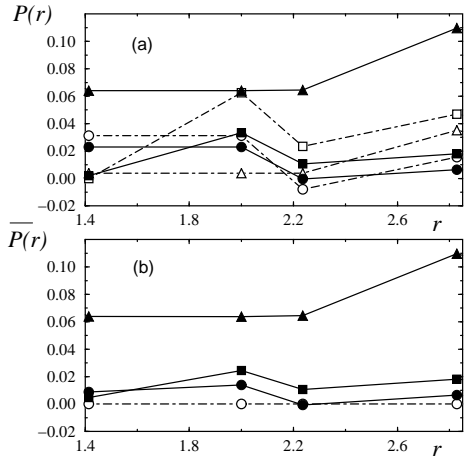,width=6cm,angle=-90}
\narrowtext
\vspace{0.4cm}
\caption{(a) Pairing correlation functions and (b) vertex contributions
to them as functions of distance for 
$U=0$ (open symbols) and $U=-5$ (solid symbols), with
$t_{AB}=1$, 
$t^{\prime }=0$,  $N=10$ and PBC.
Triangles, circles and squares correspond to on-site $s$,
extended $s$ and $d_{x^2-y^2}$ PCF respectively. Open symbols 
coincide in (b).}
\label{fig1}
\end{figure}

\begin{figure}[t]
\epsfig{file=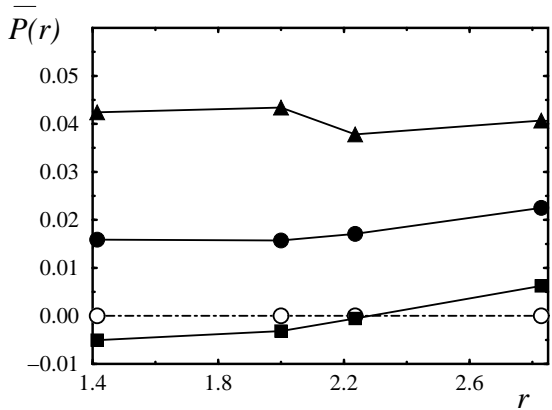,width=6cm,angle=-90}
\narrowtext
\caption{Vertex contribution to the PCF for $t_{AB}=2$, $t^{\prime}=-0.45t$, 
$U=0$, $N=10$ and PBC. The values without correlated hopping 
($t_{AB}=1$) are also shown for comparison. 
The meaning of the different symbols is the same as in Fig. 1.}
\label{fig2}
\end{figure}

\begin{figure}[htb]
\epsfig{file=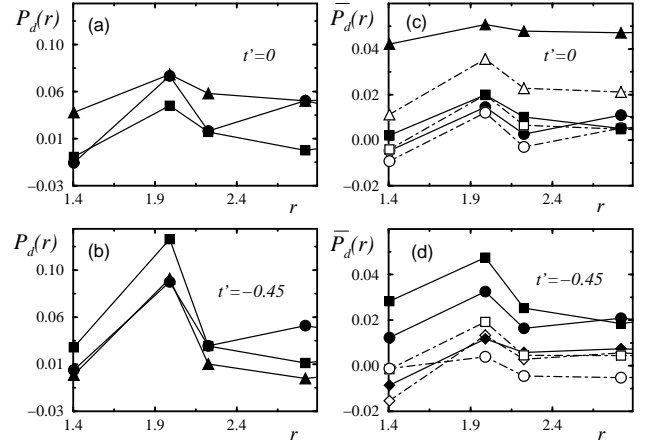,width=6cm,angle=-90}
\narrowtext
\vspace*{0.5cm} 
\caption{$P_d(r)$ and $\bar{P}_d(r)$ for $U=10$ and
$t_{AB}=1$ (open symbols) and $t_{AB}=2$ (solid symbols). 
Circles, squares and triangles correspond to $N=10, 12, 14$ 
respectively.}
\end{figure}

\newpage

\begin{figure}[htb]
\epsfig{file=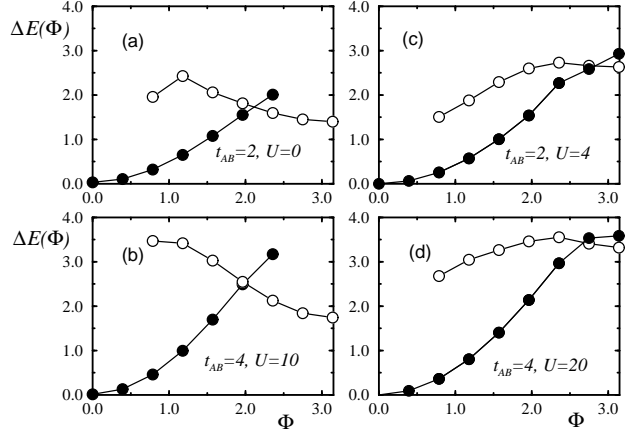,width=6cm,angle=-90}
\narrowtext
\vspace{0.4cm}
\caption{Lowest relevant energies as a function of flux for 
$t^{\prime }=-0.45t$, $N=10$
and several values of $U$ and $t_{AB}$. The value
$E(0)$ is subtracted. 
Only half of the interval $[0,2\pi )$
is shown because $E(\Phi )=E(-\Phi )=E(2\pi -\Phi )$. }
\label{fig4}
\end{figure}

\begin{figure}[htb]
\epsfig{file=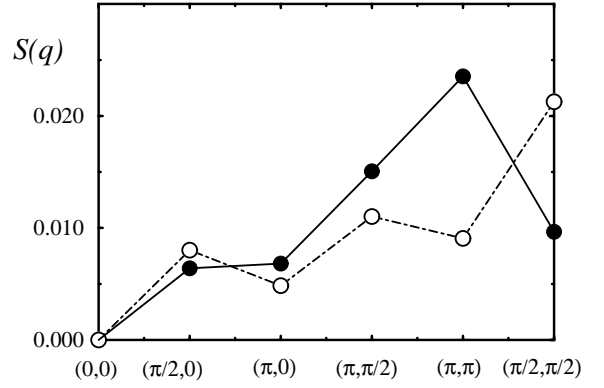,width=6cm,angle=-90}
\narrowtext
\vspace{0.4cm}
\caption{Spin structure factor as a function of wave vector for 
$t^{\prime }=-0.45t$, $U=10$, $N=12$.
Open (solid) circles denote $t_{AB}=1 (t_{AB}=2)$.
}
\label{fig5}
\end{figure}

\end{multicols}


\begin{references}
\bibitem{note}  When comparing with other works, note that our normalized 
$P_{es}(i)$ and $P_{d}(i)$ are eigth times smaller than previous 
definitions \cite{rvb,zha}.
\bibitem{notebc} For $t'=0$, $N=14$, the OBC are MBC. As a consequence the 
ground state is highly degenerate in the non-interacting limit, and this 
can affect the behavoir of the PCF.
\end{references}
\end{document}